# Data-Driven Dystopia: an uninterrupted breach of ethics

Shreyansh Padarha
CHRIST (Deemed to be University), Pune Lavasa Campus

"A better world won't come about simply because we use data; data has its dark underside." - Mike Loukides
[1]

**ABSTRACT**

This article discusses the risks and complexities associated with the exponential rise in data and the misuse of data by large corporations. The article presents instances of data breaches and data harvesting practices that violate user privacy. It also explores the concept of "Weapons Of Math Destruction" (WMDs), which refers to big data models that perpetuate inequality and discrimination. The article highlights the need for companies to take responsibility for safeguarding user information and the ethical use of data models, AI, and ML. The article also emphasizes the significance of data privacy for individuals in their daily lives and the need for a more conscious and responsible approach towards data management.

## [1] Introduction

Around 150 years back, Charles Babbage, the inventor of the first digital computer, proclaimed "Errors using inadequate data are much less than those using no data at all." [2]. This quote has stuck through the test of time, data is what surrounds humans. In 2020, when the world was locked behind doors, on average we created a mammoth 2.5 quintillion bytes of data daily [3]. The exponential rise in data has brought along endless opportunities and possibilities. But the complexities and risks involved are in abundance as well. Nowadays it is typical for a data breach, theft or misuse report to pop up in our news feed. From the likes of JP Morgan Chase, Sony, Home Depot, and Anthem to the juggernaut, that is The White House in DC, all have been victims of data breaches [4]. Then there are "the usual suspects", Google, Facebook, and Amazon infamous and held guilty [EU] for continuous abuse of power, privacy violations and misusing private consumer data on numerous occasions [5] [6] [7].

## [2] Negligence in handling "consented" customer data – Data Breaches

Yahoo in 2016, came into the spotlight for two separate instances of security breaches. In 2013, 500 million accounts within Yahoo were hacked, and in 2014 another 1 billion users were compromised [8]. The consequences of which are untraceable yet significant. To make things worse, in October of 2017, Yahoo confirmed that 3 billion of its user accounts were impacted. The hackers had access to all cookies, which meant the password, name, dob, security questions, address and telephone number of each account were accessible to them, causing the biggest data breach in the history of the Internet [9].

Such an instance is not a standalone one, in march 2021, a hacker forum published leaked private Facebook data of 533 million people, from across 106 countries [10]. Although Facebook claims the data was from 2019 and they are not guilty of any charges, the repercussions and consequences are not bound by time [11] [12].

It's highly likely that while agreeing to the terms and conditions on a website, or while creating an account, we won't be reading the consent form. Despite the many rights the user is stripped of when accepting the terms, the user does not subscribe for their data being stolen or hacked into. It's the company's responsibility to safeguard the user's information [13].

## [3] Reckless misuse of sensitive data by large firms

Although data breaches, thefts and hacking into servers are malpractices, where the companies responsible for user data should be held accountable, the corporations don't have malice or ill will towards the users [14]. But, practices like data harvesting, data tracking and third-party selling have been exercised even after knowing the possible adverse consequences [15] [16]. As depicted in the table below, all major FAANG companies have been involved in such privacy violations [17].

| S. No. | Company | Misconduct (description) |
|---|---|---|
| 1. | **Twitter** | Twitter, on 17 September 2019, admitted that the credential details provided by their users "may have inadvertently been used for advertising purposes", as quoted from the official website [18]. |
| 2. | **Amazon** | The European commission currently has an open investigation running on Amazon for possible breach of competition rules, where they've allegedly been using sensitive data from independent retailers to improve sales of Amazon produced products [19]. |
| 3. | **Facebook (Cambridge Analytica)** | The biggest example of data misuse in recent history is the British firm Cambridge Analytica, acquiring and misusing data of 87 million Facebook users, to aid the 2016 Presidential Campaign of Republican Candidate Donald Trump [20] [21]. All of this came to light through a whistle-blower in 2018 [22]. |

**Table 1. Company-wise Data Breach**

## [4] WMDs

Since the Cold War, WMDs or Weapons Of Mass Destruction have been discussed in various security council meetings. It's a topic profound for provoking debates between countries [23]. But in the contemporary world, WMDs, have gained a new meaning. The American mathematician Cathy O'Neil, has coined a new term (her book's title), "Weapons Of Math Destruction" [24]. O'Neil argues that Big data and data models based on maths have increased inequality and threatened democracy. She also asserts that WMDs (big data models) have consequences for everyone, but the aftermaths are intractable for minorities [25].

Computer systems have been the epicentre of all forms of discrimination. The question has been raised multiple times, How do we know A.I. machines aren't racist? The Compas (Correctional Offender Management Profiling for Alternative Sanctions) system, used by US courts for risk assessment, was much more likely to mistakenly categorise black defendants as "likely to re-offend" [26]. Even in interview screenings, the CVs of coloured individuals have been rejected based on current employee databases of companies [27].

It's worth pondering over, whether big data models, ML, A.I., are being misused and are becoming a medium of unethical practices.

## [5] Conclusion

All the instances mentioned above of data being misused or appalling mismanagement of personal data, whether they were Intentional or unintentional the consequences were borne by the "common man". Our daily lives have been entangled with data, even the barest of tasks, like taking an Uber taxi to the airport have been sources of exploitation of private data. Uber, with its "god-view", had the access to real-time location of customers and drivers, crossing multiple privacy boundaries [28]. It is imperative to hold these companies accountable for their wrongdoings and to have tighter regulations imposed on them.

# REFERENCES

[1] H. M. a. D. P. Mike Loukides, Ethics and Data Science, O'Reilly Media, 2018.

[2] C. Babbage, apx. 1856.

[3] B. Marr, "How Much Data Do We Create Every Day? The Mind-Blowing Stats Everyone Should Read," 21 May 2018 . [Online]. Available: https://www.forbes.com/sites/bernardmarr/2018/05/21/how-much-data-do-we-create-every-day-the-mind-blowing-stats-everyone-should-read/?sh=36f202ea60ba .

[4] E. Nakashima, "Hackers breach some White House computers," 28 October 2014. [Online]. Available: https://www.washingtonpost.com/world/national-security/hackers-breach-some-white-house-computers/2014/10/28/2ddf2fa0-5ef7-11e4-91f7-5d89b5e8c251_story.html.

[5] J. Cox, "Leaked Document Says Google Fired Dozens of Employees for Data Misuse," 4 August 2021. [Online]. Available: https://www.vice.com/en/article/g5gk73/google-fired-dozens-for-data-misuse.

[6] N. Lomas, "Facebook data misuse and voter manipulation back in the frame with latest Cambridge Analytica leaks," 6 January 2020. [Online]. Available: https://techcrunch.com/2020/01/06/facebook-data-misuse-and-voter-manipulation-back-in-the-frame-with-latest-cambridge-analytica-leaks/ .

[7] W. Evans, "Amazon's Dark Secret: It Has Failed to Protect Your Data," 18 November 2021. [Online]. Available: https://www.wired.com/story/amazon-failed-to-protect-your-data-investigation/.

[8] N. P. Vindu Goel, "Yahoo Says 1 Billion User Accounts Were Hacked," 14 December 2014. [Online]. Available: https://www.nytimes.com/2016/12/14/technology/yahoo-hack.html.

[9] R. M. a. R. Knutson, "Yahoo Triples Estimate of Breached Accounts to 3 Billion," 13 October 2017. [Online]. Available: https://www.wsj.com/articles/yahoo-triples-estimate-of-breached-accounts-to-3-billion-1507062804.

[10] A. Holmes, "533 million Facebook users' phone numbers and personal data have been leaked online," 4 April 2021. [Online]. Available: https://www.businessinsider.in/tech/news/533-million-facebook-users-phone-numbers-and-personal-data-have-been-leaked-online/articleshow/81889315.cms.

[11] BBC, "Facebook downplays data breach in internal email," 20 April 2021. [Online]. Available: https://www.bbc.com/news/technology-56815478#:~:text=Data%20from%20533%20million%20people,available%20information%20on%20the%20site.

[12] E. Bowman, "After Data Breach Exposes 530 Million, Facebook Says It Will Not Notify Users," 9 April 2021. [Online]. Available: https://www.npr.org/2021/04/09/986005820/after-data-breach-exposes-530-million-facebook-says-it-will-not-notify-users.

[13] R. W. B. Z. Leon Trakman, "Digital Consent and Data Protection Law – Europe and Asia-Pacific Experience," *SSRN,* p. 33, 2020.

[14] B. NeSmith, "CEOs: The Data Breach Is Your Fault," 26 June 2018. [Online]. Available: https://www.forbes.com/sites/forbestechcouncil/2018/06/26/ceos-the-data-breach-is-your-fault/?sh=4de2b57458b0.

[15] Reuters, "Big Tech data harvesting comes under fire by world central bank group," 06 May 2022. [Online]. Available: https://economictimes.indiatimes.com/tech/technology/big-tech-data-harvesting-comes-under-fire-by-world-central-bank-group/articleshow/91363642.cms.

[16] S. Morrow, "Data breach vs. data misuse: Reducing business risk with good data tracking," 9 November 2020. [Online]. Available: https://resources.infosecinstitute.com/topic/data-breach-vs-data-misuse-reducing-business-risk-with-good-data-tracking/.

[17] dictionary.com, *https://www.dictionary.com/e/acronyms/faang/.*

[18] "Twitter Help Center," [Online]. Available: https://help.twitter.com/en/information-and-ads.

[19] P. r. EC, "Antitrust: Commission opens investigation into possible anti-competitive conduct of Amazon," 17 July 2019. [Online]. Available: https://ec.europa.eu/commission/presscorner/detail/en/IP_19_4291.

[20] N. Confessore, "Cambridge Analytica and Facebook: The Scandal and the Fallout So Far," 4 April 2018. [Online]. Available: https://www.nytimes.com/2018/04/04/us/politics/cambridge-analytica-scandal-fallout.html.

[21] R. Nieva, "Most Facebook users hit by Cambridge Analytica scandal are Californians," 13 June 2018. [Online]. Available: https://www.cnet.com/tech/tech-industry/most-facebook-users-hit-by-cambridge-analytica-scandal-are-californians/.

[22] T. Gross, "Whistleblower Explains How Cambridge Analytica Helped Fuel U.S. 'Insurgency'," 8 October 2019. [Online]. Available: https://www.npr.org/2019/10/08/768216311/whistleblower-explains-how-cambridge-analytica-helped-fuel-u-s-insurgency.

[23] US Department of State, "Remarks on United Nations Security Council Resolution 1540," 23 May 2022. [Online]. Available: https://www.state.gov/remarks-on-united-nations-security-council-resolution-1540/.

[24] C. O'Neil, Weapons of Math Destruction, Penguin Books Limited, 2016.

[25] T. Woodson, "Weapons of math destruction," *Journal of Responsible Innovation,* pp. 361-363, 2018.

[26] S. Buranyi, " Rise of the racist robots – how AI is learning all our worst impulses," 8 August 2017. [Online]. Available: https://www.theguardian.com/inequality/2017/aug/08/rise-of-the-racist-robots-how-ai-is-learning-all-our-worst-impulses.

[27] C. Metz, "Who Is Making Sure the A.I. Machines Aren't Racist?," 15 March 2021. [Online]. Available: https://www.nytimes.com/2021/03/15/technology/artificial-intelligence-google-bias.html.

[28] B. Fung, "Uber settles with FTC over 'God View' and some other privacy issues," August 15 2017. [Online]. Available: https://www.latimes.com/business/technology/la-fi-tn-uber-ftc-20170815-story.html.